\RequirePackage{fixltx2e}
\documentclass[%
 aps,
 prl,%
 amsmath,amssymb,superscriptaddress,
 preprint,%
]{revtex4-1}

\usepackage{graphicx}% Include figure files
\usepackage{bm}% bold math
\usepackage{hyperref}
\usepackage{xcolor} 
\usepackage[normalem]{ulem}

\begin{document}
%opening
\title{Functionalization mediates heat transport in graphene nanoflakes}

\author{Haoxue Han}
% \email[]{haoxue.han@ecp.fr}
\affiliation{CNRS, UPR 288 Laboratoire d'Energ\'etique Mol\'eculaire et Macroscopique, Combustion (EM2C), Grande Voie des Vignes, 92295 Ch\^atenay-Malabry, France}
\affiliation{Ecole Centrale Paris, Grande Voie des Vignes, 92295 Ch\^atenay-Malabry, France}%

\author{Yong Zhang}
\affiliation{BioNano Systems Laboratory, Department of Microtechnology and Nanoscience,
Chalmers University of Technology, SE-41296 Gothenburg, Sweden}

\author{Zainelabideen Y. Mijbil}
\affiliation{Quantum Technology Center, Physics Department, Lancaster University, LA1 4YB Lancaster, UK}
\affiliation{Al-Qasim Green University, Babylon, Iraq}

\author{Hatef Sadeghi}
\affiliation{Quantum Technology Center, Physics Department, Lancaster University, LA1 4YB Lancaster, UK}

\author{Yuxiang Ni}
\affiliation{Department of Mechanical Engineering, University of Minnesota, 111 Church Street SE, Minneapolis, MN 55455, USA}

\author{Shiyun Xiong}
\affiliation{Max Planck Institute for Polymer Research, Ackermannweg 10, D-55128 Mainz, Germany}

\author{Kimmo S\"a\"askilahti}
\affiliation{Department of Biomedical Engineering and Computational Science, Aalto University, FI-00076 Aalto, Finland}

\author{Steven Bailey}
\affiliation{Quantum Technology Center, Physics Department, Lancaster University, LA1 4YB Lancaster, UK}

\author{Yuriy A. Kosevich}%
\affiliation{CNRS, UPR 288 Laboratoire d'Energ\'etique Mol\'eculaire et Macroscopique, Combustion (EM2C), Grande Voie des Vignes, 92295 Ch\^atenay-Malabry, France}
\affiliation{Ecole Centrale Paris, Grande Voie des Vignes, 92295 Ch\^atenay-Malabry, France}%
\affiliation{Semenov Institute of Chemical Physics, Russian Academy of Sciences, Kosygin str. 4, Moscow 119991, Russia
}%

\author{Johan Liu}
\email[]{johan.liu@chalmers.se}
\affiliation{BioNano Systems Laboratory, Department of Microtechnology and Nanoscience,
Chalmers University of Technology, SE-41296 Gothenburg, Sweden}

\author{Colin J. Lambert}
\email[]{c.lambert@lancaster.ac.uk}
\affiliation{Quantum Technology Center, Physics Department, Lancaster University, LA1 4YB Lancaster, UK}

\author{Sebastian Volz}
\email[]{sebastian.volz@ecp.fr}
\affiliation{CNRS, UPR 288 Laboratoire d'Energ\'etique Mol\'eculaire et Macroscopique, Combustion (EM2C), Grande Voie des Vignes, 92295 Ch\^atenay-Malabry, France}
\affiliation{Ecole Centrale Paris, Grande Voie des Vignes, 92295 Ch\^atenay-Malabry, France}%

\newcommand{\redt}[1]{\textcolor{red}{#1}}
\date{\today}% It is always \today, today,
             %  but any date may be explicitly specified

\begin{abstract}
Self-heating is a severe problem for high-power microelectronic devices. 
Graphene and few-layer graphene have attracted tremendous attention for heat removal thanks to their extraordinarily
high in-plane thermal conductivity. However, this high thermal conductivity undergoes severe degradations
caused by the contact with the substrate and the functionalization-induced point defects.  
Here we show that thermal management of a micro heater can be substantially improved via introduction of
alternative heat-escaping channels implemented with graphene-based film covalently bonded to functionalized graphene
oxide through  silane molecules.  Theoretical and experimental results demonstrate a counter-intuitive 
enhancement of the thermal conductivity of such a graphene-based film. This increase in the in-plane 
thermal conductivity of supported graphene is accompanied by an improvement on the graphene-substrate’s
thermal contact. Using infrared thermal imaging, we demonstrate that the temperature of the hotspots 
can be lowered by ~12$^{\mathrm{o}}$C in transistors operating at ~130 W mm$^{-2}$ , which corresponds to half of 
an order-of-magnitude increase in the device lifetime. Ab initio and molecular dynamics simulations reveal
that the functionalization constrains the cross-plane scattering of low frequency phonons, which in turn 
enhances in-plane heat conduction of the bonded graphene film by recovering the long flexural phonon lifetime. 

\end{abstract}

\maketitle

\section{Introduction}
Anisotropic properties of two-dimensional (2D) layered materials make them promising in the application of 
next generation electronic device. Graphene and few-layer graphene (FLG) have been 
most intensely studied for thermal management,  due to their extraordinarily high in-plane thermal 
conductivity ($\kappa$) \cite{geim2007,balandin2008,balandin2011,kong2014}.  
However, in most practical applications, graphene/FLG  will be supported by and integrated with insulators, both in
un electronic circuitry and heat-spreader applications \cite{prasher2010}.
%(http://www.sciencemag.org/content/328/5975/185)
Therefore, thermal energy flow will be limited both by the in-plane thermal conductivity ($\kappa$) of the supported graphene/FLG
and by the thermal boundary resistance (R) at the graphene/FLG-substrate interface \cite{ong2011}. 
 % cite pop PHYSICAL REVIEW B 84, 075471 (2011)

Due to the exceedingly large surface-to-volume ratio, $\kappa$ is
very sensitive to the interactions with the environment. Indeed, when supported on an amorphous substrate, $\kappa$ 
of the suspended graphene decreased by almost one order of magnitude, from $\sim$4000 Wm\textsuperscript{-1}K\textsuperscript{-1}
\cite{ghosh2010}
% (Ghosh S, et al. Nat Mater 9(7):555–558.) 
to $\sim$600  Wm\textsuperscript{-1}K\textsuperscript{-1} \cite{sadeghi2013}. 
% %(cite Phonon-interface scattering in multilayer graphene on an amorphous support).
% $ \kappa $ of the suspended FLG decreases versus the layer number, while that of the supported counterpart exhibits 
% a different trend, which increases with the layer number. The  $ \kappa $ values of both configurations will finally 
% converge to the graphite thermal conductivity when the layer number is large enough.
% %(cite Phonon-interface scattering in multilayer graphene on an amorphous support).
Such a striking reduction in $\kappa$ significantly limits the thermal performance of graphene/FLG in real applications
and arises from the strong scattering of the important heat carrying 
flexural acoustic (ZA) modes \cite{lindsay2010}
% cite Flexural phonons and thermal transport in graphene
to the substrate \cite{seol2010}. 
% cite Two-Dimensional Phonon Transport in Supported Graphene
 %(cite Phonon-interface scattering in multilayer graphene on an amorphous support).
 % cite pop PHYSICAL REVIEW B 84, 075471 (2011)
More specifically, it was identified that the phonon relaxation times (RTs) of graphene ZA modes are suppressed 
when supported on a SiO$_{2}$ substrate.  
% These  studies have improved our fundamental understanding in the physics behind the problem and 
It was suggested that suitably choosing the substrate material \cite{sadeghi2013}
 %(cite Phonon-interface scattering in multilayer graphene on an amorphous support).
 and modulating its coupling to graphene \cite{ong2011}
% cite Effect of substrate modes on thermal transport in supported graphene 
may be useful to improve $\kappa$ of the supported graphene/FLG. 
%Unfortunately, no practical solution is proposed in countering the substrate effect, and the problem still remains unsolved.

% Begin Point two, interface resistance
The thermal boundary resistance (R) of a graphene/FLG-substrate interface is another limiting factor 
to their thermal performance in devices. Covalent functionalization has been proven to efficiently promote heat transfer
between interfaces by introducing additional thermal pathways through the functionalizing molecules 
\cite{ge2006,wang2006,ramanathan2008,konatham2009,collins2010, liang2011,chien2011,kim2012,ni2012,luo2012, hopkins2012,harikrishna2013, liang2013,obrien2013,taphouse2014,sun2014}.
For example, self-assembled-monolayers (SAM) were used to functionalize metallic surfaces to enhance heat transport across 
metal-water \cite{ge2006,harikrishna2013}, metal-gas \cite{liang2013}, metal-semiconductor \cite{wang2006} and
metal-polymer \cite{sun2014} interface.
Functionalization was used in graphene and CNT nanocomposites to mitigate the high thermal boundary resistance
between the graphene/CNT fillers and the polymer matrices \cite{liang2011, ni2012,luo2012,taphouse2014}. 
% Similar resistance reduction was observed through molecular dynamics simulations in a graphene-oil suspension where 
% the graphene flakes were functionalized with the oil molecules \cite{konatham2009}.
Functionalized molecules help align and densely pack multilayer graphene sheets and reduce the interlayer
thermal resistance of graphene \cite{liang2011}. 
Recently, it was shown that plasma-functionalized graphene was shown to increase the cross-plane thermal conductance between aluminum and its substrate
by a factor of two \cite{hopkins2012}. 
Nevertheless,  the functionalization-introduced point defects will further decrease $\kappa$ of the supported graphene/FLG, 
as they introduce phonons scattering centers \cite{chien2011,kim2012,liang2011}.

To make progress, a robust solution that maintains the high thermal conductivity of graphene/FLG when supported,
while effectively reducing the thermal interface resistance is needed. 
In this study, we theoretically and experimentally investigate the basal-plane thermal 
conductivity and membrane-substrate thermal resistance of graphene film covalently linked to a silane-functionalized 
graphene oxide (FGO) support (figure \ref{fig1}). 
Intuitively, we might expect that attaching a molecule to a graphene surface would increase phonon scattering and 
reduce $\kappa$.
We report here a counter-intuitive increase of $ \kappa $ in the supported graphene films 
with silane functionalization.
% EXP.
% Our experiment shows that the heat-spreading performance of our proposed graphene film-silane-FGO structure is decreased compared with the one without silane bonding.
We attribute this unexpected phenomenon to the silane functionalization, which mediates the graphene film-FGO substrate 
scattering and recovers the long graphene ZA modes relaxation time. 
% begin interface
Our model demonstrates that with a large number density $\rho$ of molecules, 
the overall in-plane thermal conductivity $\kappa$ of the graphene-based film is enhanced by $\sim$ 15\% with the 
presence of the FGO substrate, whereas its thermal resistance $R$ with the substrate falls below that without 
the functionalization. 
Both phenomena indicate that graphene film-silane-FGO structures is a useful tool for optimizing thermal management.

\section{Result and Discussion}
\begin{figure}
\includegraphics[width=8.8cm]{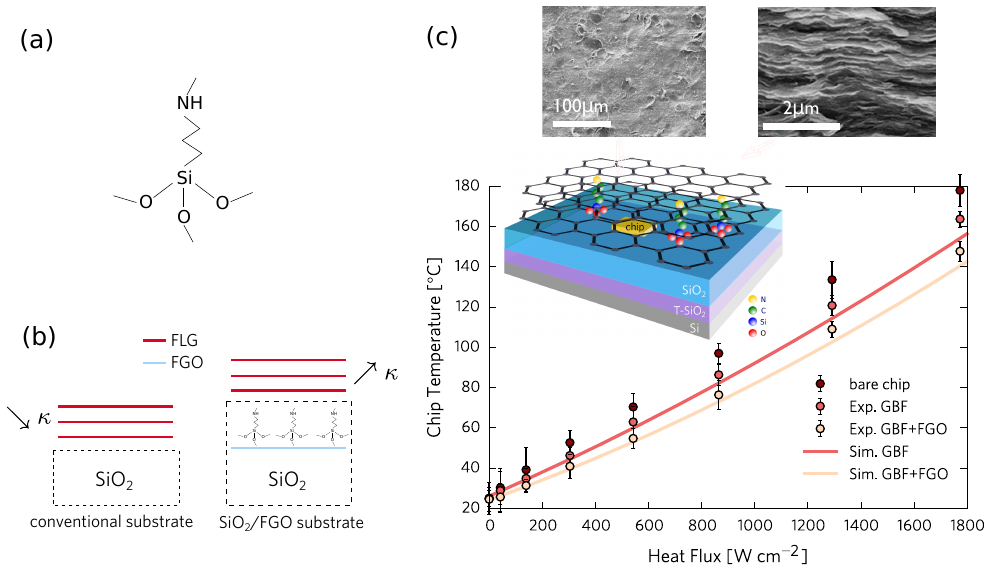}
\caption{\label{fig1}
% \textbf{\color{red} this need to be produced with better resolution}(a) Sketch of the chemical bonds of the silane molecule. 
(a) Sketch of the chemical bonds of the silane molecule.
(b) Schematic of a graphene film on different supports. Left: Conventional silica substrate. 
Right: the proposed silica / functionalized graphene oxide substrate.
(c) Measured (filled circles) and FEM simulated (lines) chip temperatures versus the in-plane heat fluxes dissipated 
in a bare chip, a chip covered by a GBF and
a chip covered by a GBF with functionalized GO. (Inset) Schematic of the measurement setup. The chip is embedded in the SiO$_2$ 
substrate. T-SiO$_2$ stands for thermally-grown SiO$_2$. 
Scanning electron microscopy (SEM) image of the in-plane and the cross section of the GBF.
}
\end{figure}

A graphene membrane bonded to the FGO substrate through silane molecules is shown in Figure \ref{fig1}(a) and (b).
To synthesize experimentally a graphene-based film and FGO (figure \ref{fig1}c), we first prepared a graphene oxide (GO) dispersion 
(See Experimental Section). The FGO was obtained by functionalizing GO with a silane-based chemistry suitable for 
reactive oxide-forming 
surfaces including the basal plane of GO and SiO\textsubscript{2}. 
Silane is widely used as a coupling agent to improve the adhesion performance of the interfaces.
The FGO layer has a thickness of $\sim$ 5 nm.
The graphene film was obtained after chemical reduction of the GO. 
The graphene film was then spin-coated \cite{pastern2008} with the FGO and the resulting bundle was transferred
to a thermal evaluation device \cite{gao2013}, resulting in the formation of molecular bridges between the graphene
surface and the device's SiO\textsubscript{2} substrate. 
The thermal evaluation device was integrated with micro Pt-based heating resistors as the hot spot and temperature
sensors \cite{gao2013}, acting as a simulation platform of an electronic component to demonstrate the
heat-spreading capability of the supported graphene film. 
Fig. \ref{fig1}(c) shows the temperature measured in situ at the hot spot and compares the thermal performance
of the graphene-film with and without the functionalization. 
With a constant heat flux of 1300 W cm\textsuperscript{-2} at the Pt hot spot, 
the temperature at the hot spot decreased from $\sim146^o$C to $\sim140\textsuperscript{o}$C 
($\Delta T\approx 6\textsuperscript{o}$C) 
with the graphene film only as heat spreader. 
In contrast, by chemically bonding the graphene film to the FGO substrate, the hot spot was cooled down
from $\sim146\textsuperscript{o}$C to $\sim134\textsuperscript{o}$C ($\Delta T\approx 12\textsuperscript{o}$C). 
The heat-spreading performance is thus enhanced by $\sim$50\%.
We have implemented a finite-element (FEM) model (See Section V in \cite{sm}) of the heat spreading device
by taking the results of atomistic simulation as input parameters. 
As shown in Figure~\ref{fig1}(c), the heat-spreading performance of the equivalent 
macroscopic FEM model agrees reasonably well with the one measured by experiments.

To explore the physics behind this counter-intuitive increase of $\kappa$ in the
supported graphene film with silane functionalization, we first performed molecular dynamic (MD) simulations to study a nanoscale 
molecular junction between two stacks of
multi-layer graphene nanoflakes. 
% A sketch of the chemical bonds of the silane molecule and the 
% We replaced the graphene oxide by pristine graphene for the simplification of computation. 
% This simplification is relevant since the oxidation of substrate has a minor impact on the
% in-plane and cross-plane thermal transport in the supported graphene membrane which we are interested in. 
We replaced the graphene oxide by pristine graphene for the simplication of computation. This simplication is relevant 
since the oxidation of substrate has a minor impact on the in-plane and cross-plane thermal transport in the supported 
graphene-based film which we are interested in.
We consider defect-free and isotopically pure graphene flakes of 10$\times$10 nm\textsuperscript{2} thermalized at 300 K 
(See methods and \cite{sm} Section I). 
The FGO substrate has a 5 nm thickness as in the experiments. 
Conventional silica substrate results in a substantial decrease of the basal-plane thermal conductivity of graphene
due to the non-conformality of the substrate-graphene interface \cite{sadeghi2013}. 
For the sake of comparison, the FGO substrate proposed herein minimizes the perturbation of substrate on the morphology of
graphene (see Fig.~\ref{fig1}(c)), thus maintaining its high thermal conductivity.
Periodic boundary conditions were used in the in-plane directions so that the MD system corresponds to two thin
films connected through silane molecules with the number density $\rho$, defined as the number of molecules per 
graphene unit area. 
% For this analysis, the current MD system of the nanoscale molecular junction will 
% help to disclose the physics of phonon transport between two graphitic nanomaterials bonded with molecules.

\begin{figure}
\includegraphics[width=7.5cm]{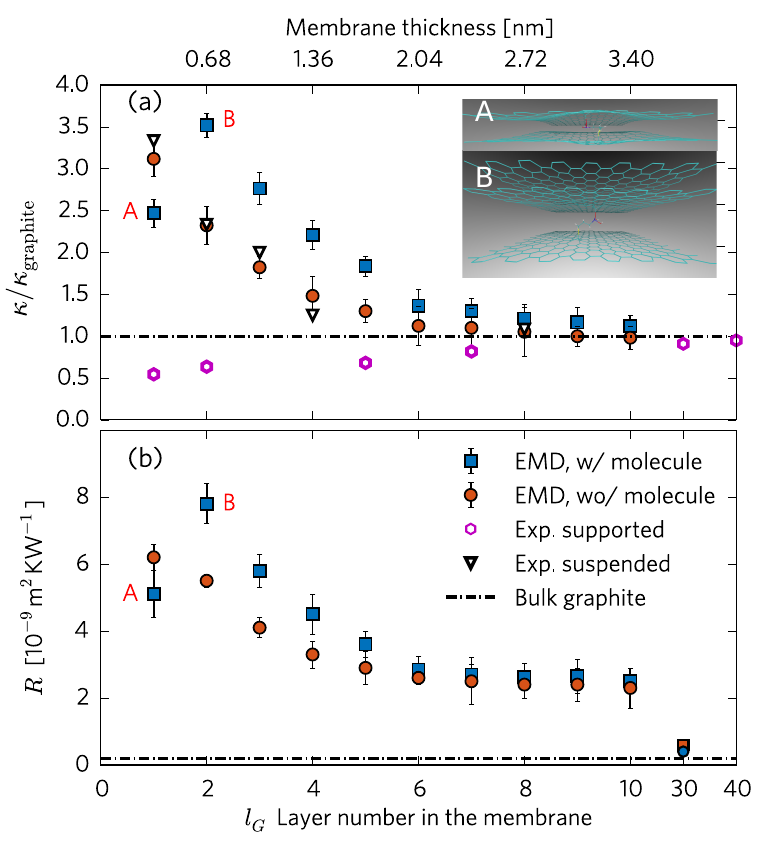}
\caption{\label{fig2}
% \textbf{\color{red} labels are not correct}
(a) MD simulation results of in-plane thermal conductivity $\kappa$ of the graphene film and 
(b) interfacial thermal resistance $R$
between the FGO substrate and the graphene film versus the graphene layer number $l_G$ in the film.
The molecule density is $\rho=0.081$ nm\textsuperscript{-2}.
Cases with (red open squres) and without (blue open circles) the molecule are compared. 
The values of thermal conductivity are normalized to that of the single-layer graphene.
Inset A illustrates a saddlelike surface generated by the molecule for $l_G=1$. Inset B shows that the 
saddlelike curvature disappears for $l_G\geq 1$. 
The measurements of suspended and supported graphene are from \cite{ghosh2010} and \cite{sadeghi2013},
respectively. The simulated (resp. measured) thermal conductivity is normalized with respect to that of the
simulated (resp. measured) graphite, in orde to allow for a reasonable comparison.
}
\end{figure}

To illustrate the intriguing role of the functionalizing molecule, 
the in-plane thermal conductivity $\kappa$ of the film and its interfacial thermal resistance $R$ \cite{sm} 
(Supplemental Information Section I)
with the FGO substrate from MD simulation combined with linear response theory
is plotted as a function of the graphene layer number $l_G$ in the film 
in Figure~\ref{fig2} (a) and (b), respectively. 
First a supercell containing a single molecule is studied, which corresponds to $\rho=0.081$ nm\textsuperscript{-2}.
For $l_G\geq 2$, the presence of the molecule results in an unexpected increase both in the graphene film thermal
conductivity $\kappa$ and in $R$. 
% \sout{The molecule-induced increase in $\kappa$ confirms the observed enhanced heat-spreading performance of 
% the graphene membrane bonded to the substrate. }
% \sout{The present simulated $\kappa$ and $R$ are in close agreement with those reported in the 
% literature for both equilibrium MD (EMD) \cite{ni2013} and non-equilibrium MD (NEMD) simulations \cite{ong2011}.
% The finite size of our simulated system may explain the notable difference between the simulated $\kappa$ and
% the measurements for supported monolayer graphene \cite{seol2010} 
% ($\sim 1100$ W~m\textsuperscript{-1}~K\textsuperscript{-1} at 300 K),
% since the current size of the graphene membrane cannot represent the phonon modes with long wavelength. 
% This discrepancy is not significant for the current work since our intention is not to present another MD
% calculation of the graphene $\kappa$, but to study the effect of the molecule coupling on the thermal transport 
% in graphene. }
An overall decaying trend of the in-plane thermal conductivity $\kappa$ of the graphene film and its
resistance $R$ 
with the substrate versus the layer number $l_G$ is observed until the value of bulk graphite is approached. 
This is due to the increased cross-plane coupling of the low-energy phonons. A similar decay was found both
in experimental measurements of $\kappa$ \cite{ghosh2010} of suspended graphene
and in simulation-based estimates of $R$ \cite{ni2013}. 
Unlike silica-supported graphene\cite{sadeghi2013}, the few-layer graphene on a functionalized graphene oxide support
recovers the high thermal conductivity and follow the same decaying trend versus $l_G$ as the suspended graphene 
\cite{ghosh2010}.
Therefore the proposed FGO substrate maintains the high thermal conductivity of graphene compared to 
a silica substrate.
For $l_G=1$, the presence of the molecule reduces $\kappa$, in contrast to the case
where no molecule interconnects the graphene film and the substrate. 
This breakdown of the thermal conductivity enhancement is due to a saddle-like surface generated around 
the molecule's chemical bonds of amino and silano groups connecting the graphene, 
with the bond center as the saddle point, as shown in the inset A of Figure~\ref{fig2} (a). 
The saddlelike surface strongly scatters the ZA phonons, thus decreasing $\kappa$ of the graphene film. 
Such a curved surface is common in defective graphene \cite{lusk2008, jiang2011}, resulting from the
Jahn-Teller effect to lower the energy by geometrical distortion \cite{hugh1983}.
% When $l_G\geq 2$, the curvature of the surface becomes imperceptible in the present system due to the 
% relatively strong adhesion of the upper graphene layers.
The thermal conductance of a single silane molecule is determined to be 82 pW K$^{-1}$ through the molecular footprint
\cite{sm} (Supplemental Information Section VI).
Our simulated result is comparable to recent measurements of the thermal conductance of alkane thiols 
SAM at a silica-gold interface\cite{meier2014}.

\begin{figure}
\includegraphics[width=8.6cm]{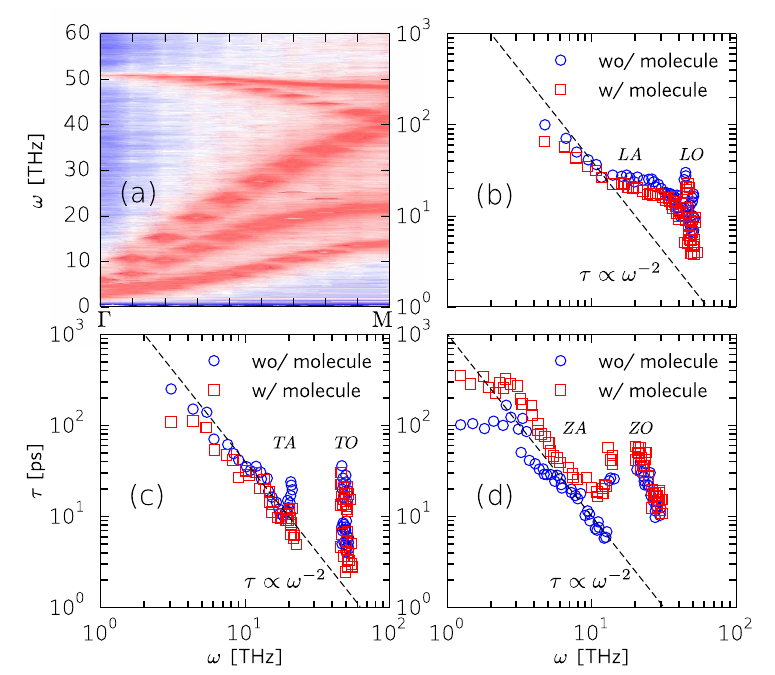}
\caption{\label{fig3}
% \textbf{\color{red} subfigure \textit{a} of this figure could go to the SI as it has not also explained in great details in the text. the y axis of the \textit{b} should be $\tau$ }
(a) Phonon dispersion of the graphene film bonded to the FGO substrate from equilibrium MD simulations, for 
$\rho=0.081$ nm\textsuperscript{-2} and $l_G$=2.
(b), (c) and (d) Mode-wise phonon relaxation time for longitudinal modes including longitudinal acoustic (LA)
and optical (LO) branches, for transverse modes including transverse acoustic (TA) and optical (TO) branches, and 
for flexural modes including flexural acoustic (ZA) and optical (ZO) branches, respectively. 
For low frequency phonons, $\tau$ follows the $\omega^{-2}$ scaling rule of the Umklapp processes at low frequency
and high-temperature limits. 
}
\end{figure}

We investigate the microscopic origin of the thermal conductivity $\kappa$ enhancement in 
the graphene film by probing the mode-wise phonon relaxation time (RT) \cite{sm} (Supplemental Information Section III). 
The phonon RT $\tau$ measures the temporal response of a perturbed phonon mode to relax back to equilibrium 
due to the net effect of different phonon scattering mechanisms. $\tau$ can be defined as \cite{ziman1960,thomas2010,qiu2012}
$\partial n/\partial\tau=(n-n_0)/\tau$ where $n$ and $n_0$ are the phonon occupation numbers out of and at
thermal equilibrium. Under the single-mode-relaxation-time approximation, the thermal conductivity is defined
by $\kappa=1/3\sum_i C_i v_i^2 \tau_i$ where $C_i$ and $v_i$ are the specific heat per volume unit and the 
group velocity of the $i$-th phonon mode. 
The phonon dispersion of the supported graphene film for $\rho=0.081$ nm\textsuperscript{-2} and $l_G$=2 
and the extracted RT $\tau$ for all the phonon modes at 300 K are shown in Figure \ref{fig3}. 
By inserting the molecule, the RT of the acoustic flexural modes $\tau_{ZA}$ largely increases at
low frequencies ($\omega<10$ THz) whereas the longitudinal and transverse modes undergo a slight decrease in 
$\tau_{LA,LO,TA,TO}$. The notable increase in $\tau_{ZA}$ accounts for the enhancement in $\kappa$ of the 
graphene film bonded to the substrate since the ZA modes contribute considerably to $\kappa$ as much as
77\% at 300 K \cite{lindsay2010}. 
We attribute the increase in $\tau_{ZA}$ to the weakened coupling between the graphene film and the substrate,
which is reflected by the increased thermal resistance $R$ for $\rho=0.081$ nm\textsuperscript{-2} 
and $l_G\geq 2$, as is shown in Fig. \ref{fig2}(b). An approximate expression from the perturbation theory
for the RT due to phonon leakage towards the contact interface yields $\tau^{-1}\propto g(\omega)K^2/\omega^2$,
where $g(\omega)$ depends on the phonon density of states and $K$ is the average van der Waals (vdW) coupling constant between
the graphene film and its substrate. A previous calculation \cite{seol2010} shows that $K_{ZA}$ largely 
depends on the substrate morphology. The presence of the molecule strongly decreases the film-substrate coupling
$K_{ZA}$, thus yielding a longer time $\tau_{ZA}$. 
This demonstrates that the silane molecules constrain the film-substrate phonon scattering, which in turn
enhances in-plane heat conduction in the bonded graphene film. 
Note that ZA mode becomes ``massive'' as shown in Fig.~\ref{fig3}(a), 
i.e., the ZA branch does not reach all the way to zero frequency but shifts to higher frequencies. 
This might reduce the thermal conductivity, but the effect is weaker than that of the 
reduced plane-plane scattering, so the in-plane thermal conductivity increases after all.

\begin{figure}
\includegraphics[width=9.2cm]{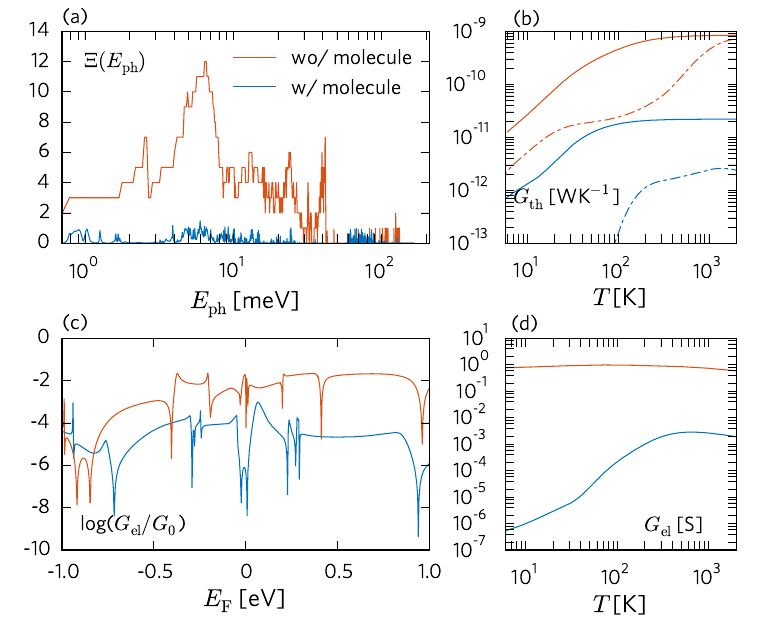}
\caption{\label{fig4}
% \textbf{\color{red} could you please replot b and d with normal scale x axis and upto 400 K. Could you also use $\kappa$ for thermal conductance? Also could you replace d with the electronic thermal conductance vs. T}
(a) Phonon transmission $\Xi(E_{\mathrm{ph}})$ versus phonon energy $E_{\mathrm{ph}}=\hbar\omega$
(red curve) between two adjacent 
graphene layers and (blue curve) through the silane molecule bonding the two graphene layers. 
(b) Thermal conductances $G_{\mathrm{th}} = G_{\mathrm{e}} +G_{\mathrm{ph}} $ contributed by phonons $G_{\mathrm{ph}}$
(solid lines) and electrons $G_{\mathrm{e}}$ (dashed lines) 
versus temperature for the two cases.
(c) Electron transmission $T_{\mathrm{el}}=G_{\mathrm{el}}/G_0$ versus Fermi energy $E_{\mathrm{F}}$ between two adjacent
graphene layers and (blue curve) through the silane molecule bonding the two graphene layers. 
(d) Electrical conductances $G_{\mathrm{el}}(T)$ versus temperature for the two cases.
}
\end{figure}

For heat transport along short molecules such as silane, the internal vibrational properties of
the molecule are crucial \cite{meier2014}. 
To gain a microscopic insight into the thermal transport at the graphene film-substrate interface 
mediated by the functionalizing molecule, 
we investigated how the differences in the phonon transmission impact the interfacial phonon transport. 
To this aim, we probed the phonon transmission $\Xi(E_{\mathrm{ph}})$ by an atomistic Green's Function
(AGF) technique\cite{sm} (Supplemental Information Section II) to characterize the local heat conduction with and 
without the presence of the molecule. 
% \redt{
$\Xi(E_{\mathrm{ph}})$ enables a precise measurement of the atomic-scale molecule-graphene heat transport
that the conventional models like AMM \cite{little1959} and DMM \cite{swartz1989} fail to provide. 
% }
As shown in Figure \ref{fig4},
the transmission function $\Xi(E_{\mathrm{ph}})$ for the two adjacent graphene layers without any molecule 
displays a clear stepwise structure that follows the number of open phonon channels. 
% In the low-energy region below 1.3 THz, being the energy gap of the lowest optical modes, the dashed curve
% shows $\Xi(\omega)=4$, indicating the number of acoustic branches corresponding to longitudinal,
% twisting, and doubly degenerated flexural modes. This latter value reflects the perfect transmission of 
% all acoustic modes. 
Low energy phonons ($E_{\mathrm{ph}}\leq 10$ meV) dominate heat conduction since
the adjacent graphene flakes interact only through weak vdW forces that inhibit the transmission of high-frequency 
phonons\cite{yang2014}. 
When the graphene layers are bridged by a silane molecule, the high-frequency phonons act as the major contributors
in the heat conductance $G_{\mathrm{ph}}$, creating more phonon channels through the covalent bond vibrations. 
This is in line with the transmission calculation of Segal \textit{et al.} \cite{segal2003} who observed 
a contribution to heat conduction by higher-frequency phonons within the molecule coupled to the low-frequency
phonons responsible for heat transport in the thermal reservoirs.
The oscillations in the transmission spectrum may originate from phonon interferences within the alkane
chain \cite{hu2010,markussen2013,han2014,han2015}. Fabry-P\'erot (FP) like interference effects occur in the frequency 
region of $E_{\mathrm{ph}}=20\sim 100$ meV, as was previously observed in an alkane SAM interface
\cite{hu2010}. Such FP like interferences originate from the multiple reflected phonons interfering
constructively within the alkane chain, as the local maxima in the transmission (Figure~\ref{fig4}) 
through the molecule can attain
the same intensity of that through pristine graphene films at given frequencies.
Although destructive quantum interference was believed unlikely to occur in a linear alkane chain
\cite{markussen2013},
we observe strong destructive interference patterns in the high frequency range ($\omega=40\sim 60$ THz),
which may correspond to two-path destructive phonon interferences \cite{han2014}. 

We investigated the inter-layer electron transport in the graphene and the effect of the silane intercalation 
in such a hybrid nanostructure through a density-functional-theory calculation combined with Green's function (Methods).
As shown in Fig.~\ref{fig4} (c) and (d), the presence of the molecule reduces the electron transmission by interrupting the $\pi-\pi$
stacking of the phenyl rings in the adjacent graphene flakes. 
This means that
the silane molecule eliminates the electrical current leakage by isolating the graphene heat spreader and the 
electronic device. 
Consequently, the main source of the thermal conductance in this system is phonons as the thermal conductance due to 
the electrons contributes to the total thermal conductance by about $20$\% (Methods).

\begin{figure}
\includegraphics[width=7.6cm]{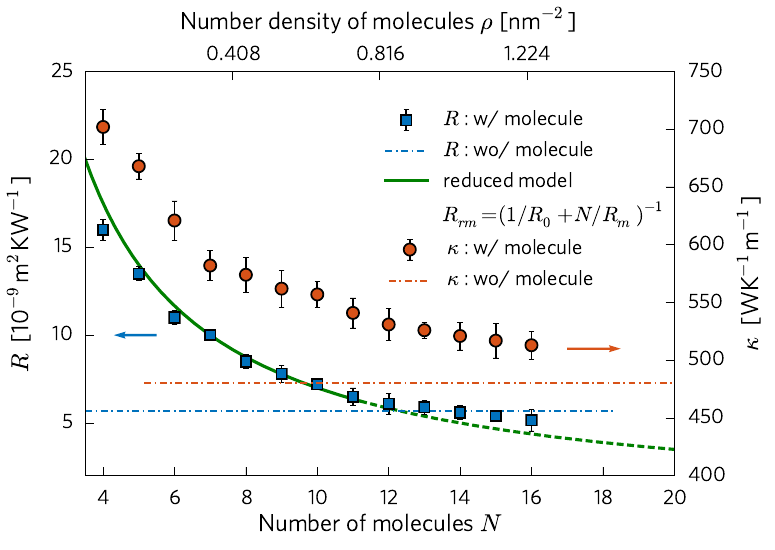}
\caption{\label{fig5}
Thermal conductivity $\kappa$ (red open circles) of the graphene film and its thermal resistance $R$ 
(blue open squares) with the functionalized substrate versus the equivalent molecule number density $\rho$. 
(Green line) Resistance $R_{rm}$ predicted by a reduced-model of parallel thermal resistors is to compare with the MD results 
for small $\rho$. $R_{rm}=(1/R_0+N/R_m)^{-1}$ where $R_0$ is the thermal resistance between two adjacent
graphene layers and $R_m$ is the additional resistance induced by a single molecule. The values of $R_0$ and 
$R_m$ are determined in MD simulations: $R_0=0.203$ mm\textsuperscript{2}KW\textsuperscript{-1} and 
$R_m=0.069$ mm\textsuperscript{2}KW\textsuperscript{-1}. 
Red and blue dashed lines represent, respectively, $\kappa$ and $R$ without the interconnecting 
molecule between at the film-substrate interface.
}
\end{figure}

We have also investigated the in-plane thermal conductivity $\kappa$ of the graphene film and 
its thermal resistance $R$ with the functionalized substrate versus the equivalent molecule number density $\rho$. 
Both $\kappa$ and $R$ decay with the number density of the molecules $\rho$.
In the limit of small number density \cite{meier2014}, we consider the molecules as independent heat conductors
connecting the film and the substrate. $R$ agrees well with a reduced-model of parallel thermal resistors. 
Thus $R$ decreases with the number of molecules. 
$\kappa$ also decreases with the molecule number since the molecules increase the cross-plane phonon scattering in the graphene film. 
For larger $\rho$, the interactions among the molecules gain importance and $R$ starts to deviate from the 
prediction of the parallel-thermal-resistor model, as shown in Figure~\ref{fig5}.
For large $\rho$, the thermal resistance $R$ is lower than that without molecules.
However $\kappa$ remains enhanced by a factor of $\approx 15$\% compared to its value for the non-functionalized graphene substrate. 
The molecule density $\rho$ effectively tunes the thermal conductivity of the supported graphene.

\section{Conclusions}
We address the long-standing problem of detrimental effect of covalent functionalization on the thermal
conductivity of supported few-layer graphene. An unprecedented increase in the basal-plane thermal conductivity of 
supported graphene was achieved while improving the graphene-substrate's thermal contact. 
Atomistic calculations show that the functionalization constrains the cross-plane scattering of 
low frequency phonons, which in turn enhances in-plane heat conduction of the bonded graphene film
through phonon lifetime prolongation.
Furthermore, we achieve a flexible tuning of the in-plane thermal conductivity of the supported graphene
by adjusting the molecule density. 
Our atomistic calculations provide a microscopic insight and disclose the physics of phonon
transport between two graphitic nanomaterials bonded with molecules. 
The detailed analysis reveals the intriguing role of molecular wires in controlling the heat transfer at the atomic-scale 
interface, which classical models fail to provide. 
Such a strategy is experimentally proven to be viable and important to the engineering of 
graphene electronics with improved heat transfer properties. 

\section{Methods}
% \textit{Sample Synthesis.} 
% Graphite (Sigma, 4g), H2SO4 (92 mL, 98$\%$), NaNO3 (2g) and KMnO\textsubscript{4} (12g) was used to prepare
% graphene oxide (GO) dispersion by following Hummer’s Method \cite{hummers1958}.
% The obtained GO dispersion was reduced by L-ascorbic acid (LAA), polyvinyl alcohol (PVA) was also added for
% better suspension. The graphene-based membrane was prepared via vacuum filtration with polycarbonate filter
% paper with the pore size of 3 $\mu$m. The membrane thickness 
% was controlled by the filtration volume and the graphene concentration in the suspension. 
% After dissolving the filter paper in pure acetone, a freestanding 
% graphene-based film was obtained. The thickness of the graphene-based membrane was measured as
% approximately 20 $\mu$m. 
% 
% \textit{Functionalization.} GO powder (20 mg) and Dicyclohexylcarbodiimide (DCC) (5 mg) were mixed with
% (3-Aminopropyl) triethoxysilane (APTES) (30 mL) by ultrasonication for 2h to produce a homogeneous suspension.
% Then the suspension was heated up to 100$^o$C for 3h with continuous stirring to realize the functionalization.
% 
% 
% \textit{Transfer process.} The graphene-based membrane was first transferred 
% onto a thermal release tape, and then spin-coated \cite{pastern2008} with the FGO layer at 4000 rpm for 2 min 
% onto the membrane. The thermal release tape was removed by heating the device. 

\textit{Simulation setup.} classical MD simulations were performed using {\footnotesize{LAMMPS}} \cite{plimpton1995}. 
The adaptive intermolecular reactive empirical bond order (AIREBO) potential \cite{stuart2000} was used to 
simulate the graphene's
C-C interactions. The long-range electromagnetic and the short-range repulsive-attractive interactions 
in the molecule is taken into account through the van Beest, Kramer, and van Santen (BKS) potential.
composed of Coulomb and Buckingham potentials. 
Periodic boundary conditions are applied in the in-plane
directions and free boundary condition in the cross-plane direction of the graphene system. 
First, each super cell was relaxed at the simulation temperature to achieve zero in-plane stress. 
Then the systems were thermalized by using a Langevin heat bath. After the thermalization, 
MD runs with durations equal to 600 ps were carried out in the microcanonical ensemble to sample the 
temperature and heat flux to be used in the calculation of the thermal resistance and thermal conductivity. 
Calculation details concerning the thermal resistance, thermal conductivity, phonon transmission and 
phonon relaxation time can be found in the Supporting Information Section III\cite{sm}.

\textit{Ab initio} calculations were carried out using the quantum chemistry DFT code {\footnotesize{SIESTA}} \citep{siesta}.
All systems were first geometrically optimized in isolation, 
with a generalized local-density approximations (LDA) within Ceperley-Alder version (CA), double-zeta polarized
basis set, 0.01 eV/\AA~ force tolerance and 250 Ry mesh cutoff. 
The relaxed atomic structures can be seen in Supplemental Figures (FIG.S3).
Electron transmission coefficients were computed using the {\footnotesize{GOLLUM}} \citep{ferrer2014gollum} code.
To produce the conductance curves in Fig.~(\ref{fig4}), the
transmission coefficient $T_{\mathrm{el}}(E_{\mathrm{F}})$ was calculated for each relaxed junction geometry, by
first obtaining the corresponding Hamiltonian and then overlapping matrices with
{\footnotesize{SIESTA}} and double-zeta polarized basis set.
To produce conductance-trace curves, the
transmission coefficient $T_{\mathrm{el}}(E_{\mathrm{F}})$ was calculated for each relaxed junction geometry
and the conductance $G_{\mathrm{el}}/G_0=T_{\mathrm{el}}(E_{\mathrm{F}})$ was obtained by evaluating 
$T_{\mathrm{el}}(E_{\mathrm{F}})$ at the Fermi energy $E_{\mathrm{F}}$.

\textit{Thermal conductance.} Thermal conductance 
$G_{\mathrm{th}} = G_{\mathrm{e}} +G_{\mathrm{ph}} $ where $G_{\mathrm{ph}}$ ($G_\mathrm{e}$) 
is the thermal conductance due to the phonons (electrons). From the Phonon transmission $\Xi(E_{\mathrm{ph}})$ the
thermal conductance due to the the phonon could be calculated as 
$G_{\mathrm{ph}}(T)=\int_0^{\infty}\mathrm{d}\omega\left(- \frac{\partial f_{\mathrm{BE}}}{\partial E} \right)\hbar\omega\Xi(\omega) /2\pi$.
where $\omega$ is the frequency. $T$ refers to the mean temperature of the system, $f_{\mathrm{BE}}$ is the Bose-Einstein phonon statistics.
$k_B$ and $\hbar$ represent the Boltzmann and the reduced
Planck’s constants, respectively.
The thermal conductance due to the electrons \cite{Sadeghi1, Sadeghi2} could be calculated from the electron 
transmission coefficient $T_{\mathrm{el}}$ as $G_{\mathrm{e}}(T)= \frac{L_0L_2-L_1^2}{hTL_0}$ where 
$L_n=\int_{-{\infty}}^{\infty}\mathrm{d}E(E-E_F)^nT_{\mathrm{el}}(E) \left(- \frac{\partial f_{\mathrm{FD}}}{\partial E} \right)$ where 
$f_{\mathrm{FD}}$ is the Fermi-Dirac electron statistics. 

\textit{Experimental method.} 
(i) Sample Synthesis.
Graphite (Sigma, 4g), H$_2$SO$_4$ (92 mL, 98$\%$), NaNO$_3$ (2g) and KMnO\textsubscript{4} (12g) was used to prepare
graphene oxide (GO) dispersion by following Hummer’s Method \cite{hummers1958}.
The obtained GO dispersion was reduced by L-ascorbic acid (LAA), polyvinyl alcohol (PVA) was also added for
better suspension. The graphene-based film was prepared via vacuum filtration with polycarbonate filter
paper with the pore size of 3 $\mu$m. The film thickness 
was controlled by the filtration volume and the graphene concentration in the suspension. 
After dissolving the filter paper in pure acetone, a freestanding 
graphene-based film was obtained. The thickness of the graphene-based film was measured as
approximately 20 $\mu$m. 
Raman spectroscopy data of the GBF before and after the spin-coating with FGO is shown in Supporting Information Section VII\cite{sm}.
(ii) Functionalization. GO powder (20 mg) and Dicyclohexylcarbodiimide (DCC) (5 mg) were mixed with
(3-Aminopropyl) triethoxysilane (APTES) (30 mL) by ultrasonication for 2h to produce a homogeneous suspension.
Then the suspension was heated up to 100$^o$C for 3h with continuous stirring to realize the functionalization.
Fourier Transform Infrared Spectroscopy (FTIR) data provides evidence for the functionalization\cite{sm}.
(iii) Transfer process. The graphene-based film was first transferred 
onto a thermal release tape, and then spin-coated \cite{pastern2008} with the FGO layer at 4000 rpm for 2 min 
onto the film. The thermal release tape was removed by heating the device.

\section{Acknowledgement}
This work was supported by the NANOTHERM project co-funded by the European Commission under the
``Information and Communication Technologies'' Seven Framework Program (FP7) and the Grant Agreement No.
318117. This work is also supported by the UK EPSRC, EP/K001507/1, EP/J014753/1, EP/H035818/1, and from the EU ITN MOLESCO 606728.

\newpage

\end{document}